\documentclass[pra,aps,twocolumn,floatfix,showpacs]{revtex4}
\usepackage{graphics}
\begin{document}
\title{Superfluid and Mott Insulating shells of bosons \\ in harmonically confined optical lattices}
\author{Kaushik Mitra, C. J. Williams and  C. A. R. S{\'a} de Melo}
\affiliation{Joint Quantum Institute \\ University of Maryland, College Park, MD 20742 \\ 
NIST, Gaithersburg, MD 20899}
\date{\today}

\begin{abstract}
Weakly interacting atomic or molecular bosons in quantum degenerate regime and trapped in harmonically
confined optical lattices, exhibit a wedding
cake structure consisting of insulating (Mott) shells. 
It is shown that superfluid regions emerge between Mott shells
as a result of fluctuations due to finite hopping.
It is found that the order parameter equation in the
superfluid regions is not of the Gross-Pitaeviskii type except near the insulator 
to superfluid boundaries. The excitation spectra
in the Mott and superfluid regions are obtained, and it is
shown that the superfluid shells posses low energy sound modes with spatially dependent sound
velocity described by a local index of refraction directly related to the local superfluid density.
 Lastly, the Berezinskii-Kosterlitz-Thouless 
transition and vortex-antivortex pairs are discussed
in thin (wide) superfluid shells (rings) limited by three (two) dimensional Mott regions. 
\end{abstract}
\maketitle

\section {Introduction}
\label{sec:introduction}

The recent experimental discovery of Bose-Mott insulating phases 
in optical lattices has generated
an explosion of research in the ultra-cold atom community (see~\cite{bloch-review-2005} for
a recent review), and has helped to merge two major branches of physics: atomic-molecular-optical and condensed
matter physics. Most experiments thus far have relied on measuring the momentum distribution of the atoms after switching off the trap confining the atoms to infer
the existence of a superfluid to insulator transition~\cite{greiner-2002, stoferle-2004, spielman-2007}. However,
very recently, two experimental groups~\cite{MIT-2006, Mainz-2006} have used spatially selective
microwave spectroscopy to probe {\it in situ} the superfluid-to-insulator transition of $^{87}$Rb 
in a three dimensional (3D) optical lattice with a harmonic envelope.
In these experiments, the shell structure of the Bose-Mott insulating states was revealed 
for very deep lattices. 
Regions of filling fraction $n = 1$ through $n = 5$ ($n = 1$ through $n = 3$) 
were mapped in the MIT~\cite{MIT-2006} (Mainz~\cite{Mainz-2006}) experiment. Their observations
in three-dimensional optical lattices lead to the confirmation of 
the Mott-insulating shell structure consisting of ``Mott plateaus'' with
abrupt transitions between any two sucessive shells as proposed in two dimensional (2D)~\cite{jaksch-1998} 
optical lattices.

One of the next frontiers for ultra-cold bosons in optical lattices is the search for
superfluid regions separating Mott-insulating shells. Eventhough the Mott shell structure 
was determined recently using microwave spectroscopy~\cite{MIT-2006, Mainz-2006}, any
evidence of superfluid shells has remained elusive. 
Thus, in anticipation of the next experimental breakthrough,
we study 2D and 3D optical lattices of atomic or molecular bosons in harmonically confining
potentials, and show that between the Mott regions of filling fraction $n$ and $n+1$,
superfluid shells emerge as a result of fluctuations due to finite hopping,
and extend our previous work on this topic~\cite{mitra-2007}.
This finite hopping breaks the local energy degeneracy of neighboring Mott-shells, 
determines the size of the superfluid regions 
and is responsible for the low energy (sound) and vortex excitations.
In addition, we find that the order parameter equation is not
in general of the Gross-Pitaeviskii type. 
Furthermore, in 3D optical lattices, when superfluid regions are thin (nearly 2D) 
spherical or ellipsoidal shells, we obtain bound vortex-antivortex excitations 
below the Berezinski-Kosterlitz-Thouless (BKT) transition temperature~\cite{berezinski, kosterlitz-thouless} 
which is different for each superfluid region. 
Finally, we propose the use of Laguerre-Gaussian and Bragg spectroscopy techniques for 
the detection of superfluid shells. 

The remainder of the manuscript is organized as follows.
In Sec.~\ref{sec:bose-hubbard-hamiltonian}, we present the
Bose-Hubbard Hamiltonian in a harmonic trap.
In Sec.~\ref{sec:shell-structure}, we analyze
the emergence of an alternating superfluid and Mott 
shell structure, comparing two different approaches
involving non-degenerate and nearly degenerate perturbation
theory. We also obtain the order parameter equation, the 
characteristic sizes of the superfluid and Mott regions, 
the local filling fraction and the local compressibility.
We find that the superfluid order parameter that emerges between
two Mott shells is not of the Gross-Pitaeviskii type, 
except very close to the insulating boundaries.
In Sec.~\ref{sec:excitations}, we describe 
quasiparticle and quasihole excitations
in the Mott regions and quasiparticle, sound and vortex 
excitations in the superfluid regions.
In Sec.~\ref{sec:experiments}, we present a possible experiment
using Gauss-Laguerre beams and Bragg spectroscopy, which 
can be performed in order to visualize the existence
of superfluid shells. Lastly, we state our conclusions in Sec.~\ref{sec:conclusions}.

\section{bose-hubbard hamiltonian}
\label{sec:bose-hubbard-hamiltonian}

To describe the physics of alternating insulating and superfluid shells
of atomic or molecular bosons in optical lattices, we use the lattice Bose-Hubbard Hamiltonian
with a harmonic potential described by
\begin{equation}
\label{eqn:bose-hubbard-hamiltonian}
H  =  -t \sum_{\mathbf{r},\mathbf{a}} c_{\mathbf{r}}^{\dagger} c_{\mathbf{r+a}}
+\frac{U}{2} \sum_{\mathbf{r}} c_{\mathbf{r}}^{\dagger} c_{\mathbf{r}}^{\dagger} 
c_{\mathbf{r}} c_{\mathbf{r}}
-\sum_{\mathbf{r}} \mu_{\mathbf{r}}
c_{\mathbf{r}}^{\dagger}c_{\mathbf{r}}
\end{equation}
where $\mu_{\mathbf{r}}= \mu - V({\bf r})$ is the local chemical potential,
\begin{equation}
\label{eqn:anisotropic-potential}
V({\bf r}) = \Omega_{\rho} (\rho/a)^2/2 + \Omega_z (z/a)^2/2
\end{equation}
is the harmonically confining potential,
$a$ is the lattice spacing, and $c_{\mathbf{r}}^{\dagger}$ is the creation operator 
for a boson at site $\mathbf{r}$. Here, the vector ${\bf r} = (x, y, z)$, and the distance
$\rho = \sqrt{x^2 + y^2}$. The anisotropic confining potential $V({\bf r})$ 
becomes isotropic when $\Omega_{\rho} = \Omega_z$.
Furthermore, $t$ is the hopping parameter and $U$ is the interaction strength, which
we assume to be repulsive (positive).
The sum in the first term on the right hand side 
of Eq.~\ref{eqn:bose-hubbard-hamiltonian} is restricted
to nearest neighbors.
The harmonic trap makes the system inhomogenous and introduces interesting
properties which are absent in homogenous Bose-Hubbard systems~\cite{fisher-1989,stoof-2001}.
For the inhomogeneous systems, it is useful to define a local chemical 
potential $\mu_{\mathbf r}=\mu-V(\mathbf r)$ which plays  
an essential role in the emergence of shell structures. 
\begin{figure}
\centerline{\scalebox{0.8}{\includegraphics{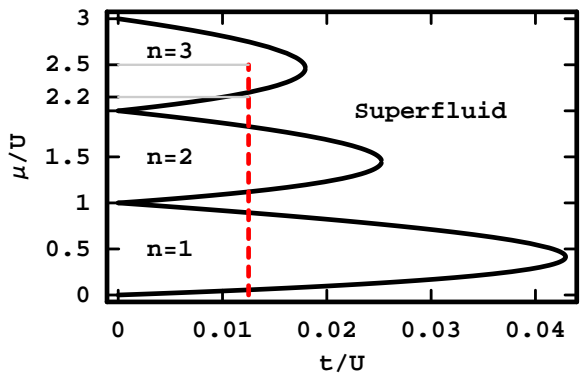}}}
\caption{\label{fig:one} (Color online) Phase diagram of the homogeneous Bose Hubbard model. 
For the inhomogeneous Bose Hubbard system, the
dashed (red) line indicates the values of the local chemical potential $\mu_{\bf r}$ that the system exhibits 
from $\mu_{\bf r}=2.5U$ (Fig~\ref{fig:1}) or $\mu_{\bf r}=2.2U$ (Fig~\ref{fig:1-2}) at the 
center of the trap to $\mu_{\bf r}=0$ at the edge of the trap for $t=1.25\times10^{-2}U$. 
It indicates how the system exhibits an alternating stucture of Mott and superfluid shells}.      
\end{figure}
For bosons in a homogeneous optical lattice, the Bose-Hubbard model without 
the harmonic confining potential can be used. In this case, it is well
known that an integer number of particles on each site or $\mathbf{r+a}$
describes an insulating state for sufficiently large $U$ ($U \gg t$). This is because 
the on-site interaction $U$ makes it energetically unfavorable
for a particle to move from one site to another. In this situation
the system is in what is known as the Mott insulator phase \cite{mott-1949}.
However, for non-integer number of particle , the extra bosons can move more 
easily, at a small energy cost, because its interaction energy is 
essentially the same on every site. For this reason,
a system with non-integer number of bosons on each site  
is a superfluid at zero temperature \cite{fisher-1989}. 
Recently, several Mott-insulator shells of bosons were detected in optical lattices~\cite{MIT-2006, Mainz-2006}, 
which are inevitably inhomogenous, and thus require a model that includes the effects of 
the harmonic part of the confining potential. 
This can be modeled by the inhomogenous Bose-Hubbard hamiltonian described
in Eq.~(\ref{eqn:bose-hubbard-hamiltonian}). In this case, one can either be in a 
regime where the entire system is superfluid or in a regime where the system 
exhibits a shell structure of alternating Mott-insulator and superfluid regions. 
The existence of this shell structure has been shown numerically~\cite{jaksch-1998},
but the lack of analytical progress has hindered a true understanding of the 
emergence and properties of these shells. 

A simple argument for the emergence of the shell structure can be
made by inspection of the standard phase diagram of the homogenous Bose-Hubbard model
shown in Fig.~\ref{fig:one} upon the substitution of $\mu \to \mu_{\bf r}$. 
In an inhomogenous system, the effective local chemical potential $\mu_{\mathbf r}$ 
is spatially varying. Thus, in the regions where the number density $n ({\bf r})$ 
is not an integer, a superfluid shell emerges, and in the regions where
$n ({\bf r})$ is an integer, a Mott-insulator shell appears.
Eventhough, a simple argument for the existence of the shell structure 
can be made, many questions need to be seriously addressed. For instance, 
what are the characteristic order parameter, dimensions and excitations of each superfluid shell? 
Thus, we begin our presentation by discussing next the emergence of the Mott-insulator
and superfluid shell structure.

\section{Emergence of the shell structure}
\label{sec:shell-structure}
A standard approach to analyse such bosonic systems is to use the
Bogoliubov mean field approximation. However, as shown in~\cite{stoof-2001},
this approximation fails to predict the expected phase transition
since it treats the interactions only approximately. Hence, instead of using the
Bogoliubov approximation, we generalize a method found in the literature~\cite{fisher-1989,stoof-2001},
by introducing a local mean field theory that treats the interactions
exactly and approximates the kinetic energy of the atoms in the optical
lattice. We introduce the local superfluid order parameter 
$\psi_{\mathbf{r}}= \langle c_{\mathbf{r}}\rangle$.
We can now construct a consistent local mean field theory by susbstituting the
operator $c_{\mathbf{r}}^{\dagger}c_{\mathbf{r+a}} \to 
\langle c_{\mathbf{r}}^{\dagger}\rangle c_{\mathbf{r+a}}
+c_{\mathbf{r}}^{\dagger}\langle c_{\mathbf{r+a}}\rangle - 
\langle c_{\mathbf{r}}^{\dagger}\rangle\langle c_{\mathbf{r+a}}\rangle$,
leading to an effective local Hamiltonian
\begin{equation}
\label{eqn:bose-hubbard-local}
H_{\mathbf{r}} ^{\textrm{eff}} = 
H_{0,n} (\mathbf{r}) 
 - t \sum_{\mathbf{a}}(c_{\mathbf{r}}\psi^{*}_{\mathbf{r+a}} 
+ c_{\mathbf{r}}^{\dagger}\psi_{\mathbf{r+a}}
-\psi^{*}_{\mathbf{r}}\psi_{\mathbf{r+a}}),
\end{equation}
which is diagonal in the site index 
$\mathbf{r}$ with 
\begin{equation}
\label{eqn:interaction-local}
H_{0,n} (\mathbf{r}) = \frac{U}{2}\hat{n}_{\mathbf{r}}(\hat{n}_{\mathbf{r}}-1)
-\mu_{\mathbf{r}} \hat{n}_{\mathbf{r}},
\end{equation}
where $\hat{n}_{\mathbf{r}} = c_{\mathbf{r}}^{\dagger}c_{\mathbf{r}}$
is the number operator. For $t = 0$, the shell structure for Mott-insulating phases 
is revealed by fixing $ \hat{n}_{\mathbf r} = n $, to obtain the local energy
\begin{equation}
\label{eqn:local-energy}
E_{0,n} (\mathbf{r}) = \frac{U}{2} n(n-1) - \mu_{\mathbf{r}} n, 
\end{equation}
when $(n - 1)U < \mu_{\mathbf{r}} < nU$. 
Since $E_{0,n+1} (\mathbf{r}) -  E_{0,n} (\mathbf{r}) = nU -  \mu_{\mathbf{r}}$,
the change from a Mott shell with filling fraction $n$ to $n + 1$ occurs
at the degeneracy condition $\mu_{\mathbf{r}} = nU$, which for a spherically symmetric potential
happens at the radius
\begin{equation}
\label{eqn:mott-shell-radius}
R_{c,n} = a \sqrt{ \Omega_n / \Omega},
\end{equation}
where $\Omega_n = 2 (\mu - nU)$.
The relation $\mu_{\mathbf{r}} = nU$ determines the shape and size
of the boundary between the $n$ and $n+1$ shells. For instance, in the
case of the anisotropic potential of Eq.~\ref{eqn:anisotropic-potential}
the same condition leads to ellipsoidal shells 
\begin{equation}
\label{eqn:ellipsoidal-mott-shell}
\left( \frac{\rho} {a_\rho} \right)^2 + \left( \frac{z} {a_z} \right)^2 = 1 
\end{equation}
with principal axes $a_\rho = a \sqrt{\Omega_n/\Omega_{\rho}}$, 
and $a_z = a \sqrt{\Omega_n/\Omega_{z}}$.  
However, near this region of degeneracy, 
fluctuations due to hopping introduce superfluid shells,
as discussed next.

\subsection{Continuum approximation}

In this section, we show how the hopping term in Eq.~(\ref{eqn:bose-hubbard-local})
affects the ground state energy of the system. Qualitatively one can
see that the kinetic energy controlled by $t$ lifts the degeneracy of 
the system at $\mu_{\mathbf{r}} = nU$ and in the process introduces 
a superfluid order parameter in a region of finite width depending on 
parameters $n,t,\Omega$ and $U$.

To obtain analytical insight into the emergence of superfluid shells, we
make first a continuum approximation through the Taylor expansion
\begin{equation}
\label{eqn:continuum}
\psi(\mathbf{r}+\mathbf{a}) = \psi(\mathbf{r})+
a_i \partial_i \psi(\mathbf{r}) + \frac{1}{2} a_i a_j \partial_i \partial_j \psi(\mathbf{r}),
\end{equation}
where repeated indices indicate summation.
This should be true in the limit where the order parameter is smoothly varying at a length scale 
much greater than 
the unit lattice spacing $a$. Under this approximation
the effective local Hamiltonian becomes 

\begin{eqnarray}
\label{eqn:bose-superfluid-local}
H_{\mathbf{r}}^{\textrm{eff}}  = 
H_{0,n} (\mathbf{r}) 
- c_{\mathbf{r}}\Delta^*(\mathbf{r}) - c_{\mathbf{r}}^{\dagger} \Delta (\mathbf{r})
+ \Lambda (\mathbf{r})
\end{eqnarray}
where $ \Delta(\mathbf{r}) =  zt\psi(\mathbf{r}) + ta^{2}\nabla^{2} \psi(\mathbf{r})  $ 
reflects the amplitude for the creation of a single boson excitation at position $\mathbf{r}$,
while the last term 
\begin{equation}
\label{eqn:lambda}
 \Lambda (\mathbf{r}) = 
\frac{1}{2} [\Delta(\mathbf{r}) \psi^{*}(\mathbf{r}) + \Delta^{*}(\mathbf{r}) \psi(\mathbf{r})]
\end{equation}
reflects the local mean-field energy shift.
Here $z = 2d$ is the number of nearest-neighbor sites (coordination number) depending on
the lattice dimension $d$. Within this approximation, a simple analytical treatment 
of the emergence of superfluid shells is possible, as discussed next.

\subsection{Nearly Degenerate Perturbation Theory}

We focus our attention now on the Mott regions with integer boson
filling $n$ and $n+1$ and the superfluid shell that emerges 
between them. In the limit where $U \gg t$ we can restrict our Hilbert
space to number-basis states $|n\rangle$ and $|n+1\rangle$ at each
site. Any contribution of other states to the local energy will be of the order of $t^{2}/U$.
The hopping term in Eq.~(\ref{eqn:bose-hubbard-local})
affects the ground state energy of the system by removing
the local degeneracy of $E_{0,n+1} (\mathbf{r})$ 
and $E_{0,n} (\mathbf{r})$ at $\mu_{\mathbf{r}} = n U$.
To illustrate this point, we use the continuum
approximation described above and write the Hamiltonian
in Eq.~\ref{eqn:bose-superfluid-local} in the matrix form
\begin{equation}
\label{eqn:h-matrix}
H_{\mathbf{r}}^{\textrm{eff}}=\left(
\begin{array}{cc}
E_{0, n} (\mathbf{r}) + \Lambda(\mathbf{r}) & -\sqrt{n+1}\Delta(\mathbf{r}) \\
-\sqrt{n+1}\Delta^{*}(\mathbf{r})  &  E_{0, n + 1} (\mathbf{r})+\Lambda(\mathbf{r})
\end{array}
\right),
\end{equation}
where $\Delta({\bf r})$ and $\Lambda(\mathbf{r})$ is defined in Eq.~\ref{eqn:lambda}.
Notice that $ t \ne 0$ has two effects. First, it changes the
local energies $E_{0, n}(\mathbf{r})$ and $E_{0, n+1}(\mathbf{r})$ of the Mott shells $n$ and $n+1$
through $\Lambda(\mathbf{r})$. Second, it mixes the two Mott regions through the off-diagonal
term $\sqrt{n+1}\Delta (\mathbf{r})$ and its hermitian conjugate. Thus, the physics near the boundary
between the $n$ and $n+1$ Mott regions is described
by an effective local two-level system with diagonal ($\Lambda(\mathbf{r})$) and
off-diagonal ($\sqrt{n+1}\Delta (\mathbf{r})$) perturbations.

\begin{figure} 
\scalebox{0.60}{\includegraphics{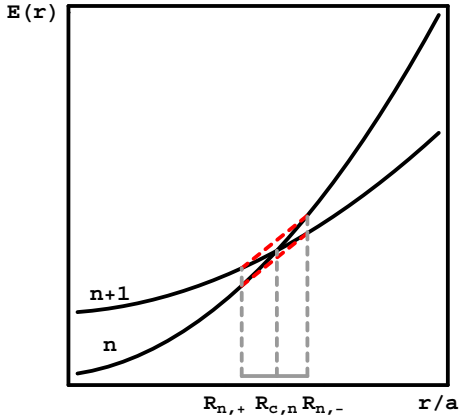} } 
\caption{\label{fig:levels}
(Color online) Schematic plot of 
local energies $E_{0,n} (\mathbf{r})$ and $E_{0,n + 1} (\mathbf{r})$ 
showing the degenerate radius $R_{c,n}$. This local energy degeneracy
is lifted by the the presence of a finite hopping $t$, which leads to 
an avoided level crossing shown as dashed dark-grey (red) curve, and to the emergence of a superfluid region 
with inner radius $R_{n, -}$ and outer radius $R_{n, +}$
}
\end{figure}

The eigenvalues of Eq.~(\ref{eqn:h-matrix}) are given by, 
\begin{equation}
\label{eqn:eigenvalues}
E_{\pm} (\mathbf{r}) 
= E_s (\mathbf{r}) \pm\sqrt{ (E_d (\mathbf{r}))^2
+ (n+1)\left|\Delta(\mathbf{r})\right|^{2}},
\end{equation}
where $E_s(\mathbf{r}) = \left[ E_{0, n+1}(\mathbf{r})+ E_{0, n}(\mathbf{r})\right]/2 + \Lambda(\mathbf{r})$
is proportional to the sum of the diagonal terms, and
$E_d(\mathbf{r}) = 
\left[ E_{0, n+1}(\mathbf{r})-  E_{0, n}(\mathbf{r})\right]/2 = ( nU - \mu_{\mathbf{r}} )/2$
is proportional to their difference. 
These local eigenvalues are illustrated schematically in Fig.~\ref{fig:levels}, 
where the local energies $E_{0,n} (\mathbf{r})$ and $E_{0,n + 1} (\mathbf{r})$ 
are shown together with the eigenvalues $E_{\pm} (\mathbf{r})$. The radii
$R_{n,\pm}$ indicated in the figure correspond to the locations where
$\Delta (\mathbf{r}) = 0$.

Notice that $E_{-} (\mathbf{r})$ is the lowest local energy, leading to 
the ground state energy 
$E = \frac{1}{L^d}\int d \mathbf{r}E_{-} (\mathbf{r}),$
where $L$ is the characteristic linear dimension of the system.

\begin{figure} 
\scalebox{0.60}{\includegraphics{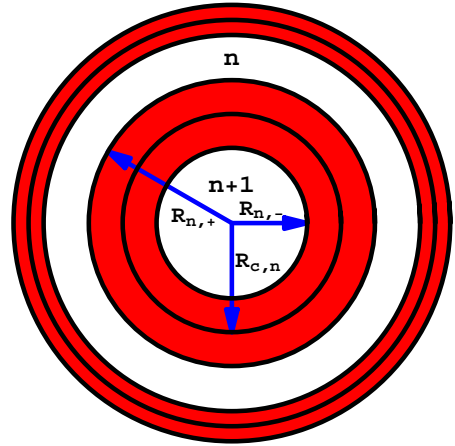} } 
\caption{\label{fig:radii}
(Color online) Schematic plot of superfluid regions, shown in gray (red),
and the Mott regions, shown in white. The radius $R_{c,n}$ separates Mott shells with filling
factor $n$ and $n+1$ for $t = 0$. The superfluid regions 
have inner radius $R_{n, -}$ and outer radius $R_{n, +}$ and 
emerge between Mott shells $n$ and $n+1$ for non-zero $t$.
}
\end{figure}

The order parameter equation (OPE) is determined 
by minimization of $E$ with respect to $\psi^{*}(\mathbf{r})$
leading to 
\begin{equation}
\label{eqn:order-parameter}
\Delta(\mathbf{r})
-\frac{(n+1) t (z+a^{2}\nabla^{2})\Delta(\mathbf{r})}
{2\sqrt{ \vert  E_d(\mathbf{r}) \vert^{2}+
(n+1)\left|\Delta(\mathbf{r})\right|^{2}}}=0.
\end{equation}
Notice that the OPE is not of the Gross-Pitaeviskii (GP) type,
since the superfluid regions emerge from local fluctuations between 
neighboring Mott shells.
The zeroth order solution of this equation with $\Delta (\mathbf{r}) = zt \psi (\mathbf{r})$ leads to the  
spatially dependent order parameter 
\begin{equation}
\label{eqn:order-parameter-saddle}
\left|\psi(\mathbf{r})\right|^{2} = \frac{n+1}{4}
-\frac{\left(nU - \mu_{\mathbf{r}}\right)^{2}}{4z^{2} t^{2}(n+1)}\label{psi}.
\end{equation}
Since $\left|\psi(\mathbf{r})\right|^{2} \ge 0$, it implies that $| n U - \mu_{\mathbf{r}}| \le (n+1) z t$ in the superfluid region. Thus 
the inner radius $R_{n,-}$ and the outer radius $R_{n, +}$ of the superfluid shell 
between the $n$ and $n + 1$ Mott regions is obtained by setting 
$\left|\psi(\mathbf{r})\right|^{2} = 0$ 
leading to 
\begin{equation}
\label{eqn:radii}
R_{n, \pm}= R_{c,n} \sqrt{ 1 \pm
\frac{2 z t (n+1)}{\Omega} \frac{a^2}{R_{c,n}^2} }.
\end{equation}
for a spherically symmetric harmonic potential $V(r) = \Omega (r/a)^2/2$, where
$R_{c,n}$ is defined above. Our notation to describe the Mott-superfluid boundaries is 
illustrated in  Fig.~\ref{fig:radii}. 

Equation~\ref{eqn:radii}
shows explicitly that $t$ splits the spatial degeneracy of the $n$ and $n+1$ Mott shells at
$r = R_{c,n}$ (or $\mu_{\mathbf r} = nU$)
by introducing a superfluid region of thickness 
\begin{equation}
\label{eqn:width-superfluid}
\Delta R_n = R_{n,+} - R_{n,-}.
\end{equation}

In the case of  non-spherical harmonic potential $V(r) = \Omega_{\rho} (\rho/a)^2/2 + \Omega_z (z/a)^2/2$
the shell regions are ellipsoidal instead of spherical.
Notice that $\Delta R_n$ depends
strongly on filling fraction $n$, the ratio $zt/\Omega$ and the chemical potential $\mu$ through $R_{c,n}$.

In addition, the local filling fraction 
\begin{equation}
n(\mathbf{r}) = -\frac{\partial E_-(\mathbf{r})}{\partial
\mu}=n+\frac{1}{2}-\frac{n U - \mu_{\mathbf{r}}} {2z t (n+1)}
\end{equation}
in the same region interpolates between $n+1$ for $r \lesssim  R_{n, -}$ and
$n$ for $r \gtrsim  R_{n, +}$, while the chemical potential $\mu$ is fixed by the total number
of particles $N = {\int d \mathbf{r} n(\mathbf{r})} $.

\begin{figure} 
\scalebox{0.60}{\includegraphics{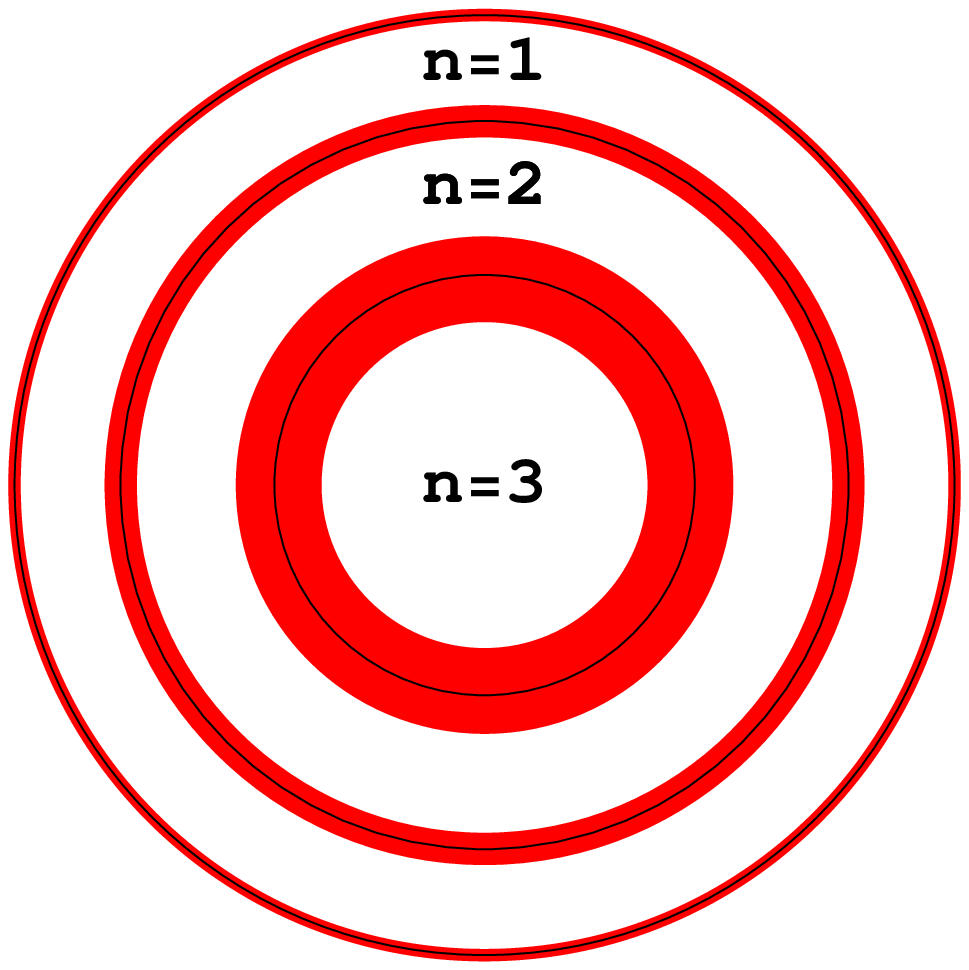}} 
\scalebox{0.70}{\includegraphics{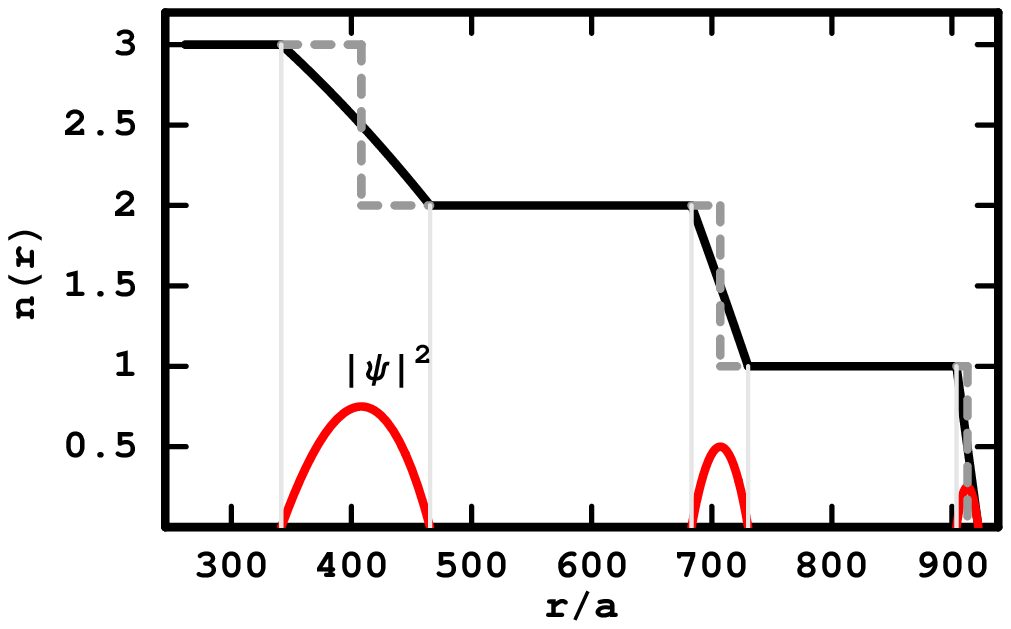} } 
\caption{ \label{fig:1}
(Color online) 
a) Shell structure of Mott and superfluids regions 
in a 2D square optical lattice with harmonic envelope 
as a function of radius $r/a$ for $t \ne 0$.
The superfluid regions are shown in red (gray)
whereas the Mott regions are shown in white. 
The black circles indicate the Mott boundaries $R_{c,n}$ at $t = 0$.
b) The local filling factor $n (\mathbf{r})$ is shown in solid lines for $t\ne 0$ 
and in dashed lined for $(t = 0)$. The red curve (solid gray) shows the local 
superfluid order parameter $|\psi ({\mathbf r})|^{2}$. 
The parameters are $\Omega= 6 \times 10^{-6} U$,
$ t = 1.25\times 10^{-2} U$ and $\mu = 2.5U$.
}
\end{figure}

\begin{figure} 
\scalebox{0.60}{\includegraphics{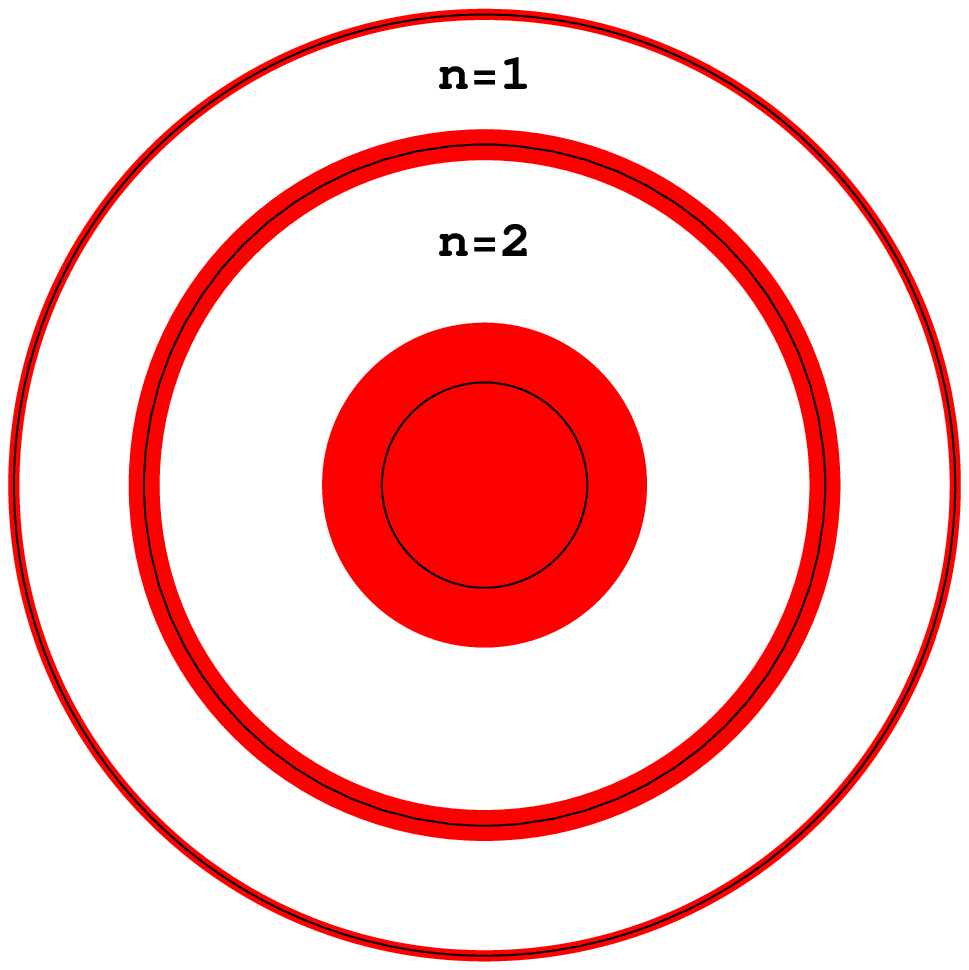} } 
\scalebox{0.70}{\includegraphics{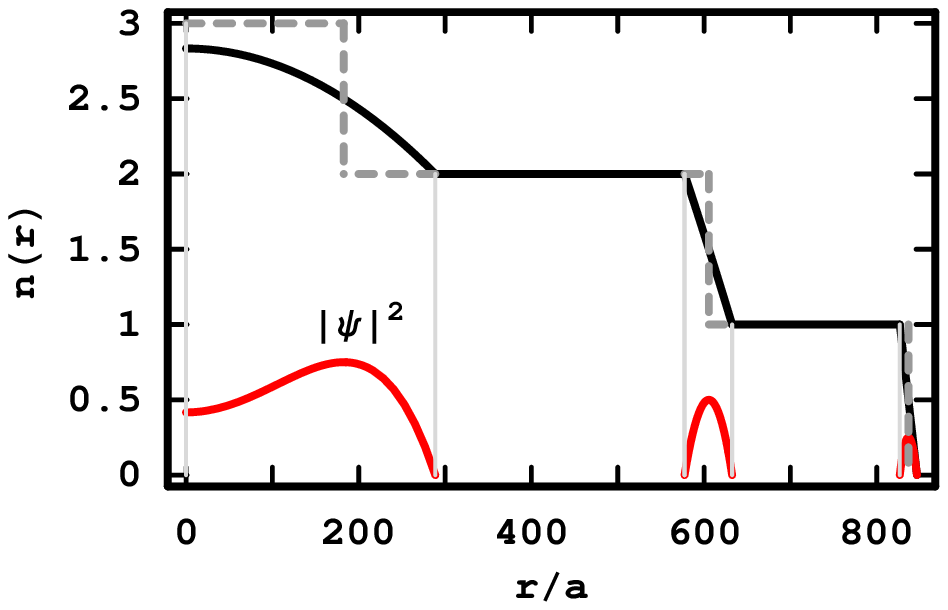} }
\caption{\label{fig:1-2}
(Color online) 
a) Shell structure of Mott and superfluids regions 
in a 2D square optical lattice with harmonic envelope 
as a function of radius $r/a$ for $t \ne 0$.
The superfluid regions are shown in red (gray)
whereas the Mott regions are shown in white. 
The black circles indicate the Mott boundaries $R_{c,n}$ at $t = 0$.
b)   The local filling factor $n (\mathbf{r})$ is shown in solid lines for $t\ne 0$ 
and in dashed lined for $(t = 0)$. The red curve (solid gray) shows the local 
superfluid order parameter $|\psi ({\mathbf r})|^{2}$. 
The parameters are $\Omega= 6 \times 10^{-6} U$,
$ t = 1.25\times 10^{-2} U$ and $\mu = 2.2U$.
}
\end{figure}

In Fig.~\ref{fig:1}, $n(\mathbf{r})$, $\left|\psi(\mathbf{r})\right|^{2}$, and
$R_{n, \pm}$ are shown for the Mott and superfluid regions, for
$t = 1.25 \times 10^{-2} U$, and $\Omega = 6 \times 10^{-6} U$, and $\mu = 2.5U$. 
For these parameters, three Mott and three superfluid shells emerge.
It is very important to emphasize that in the superfluid regions the order parameter 
$\vert \psi ({\mathbf r}) \vert^2$ is not identical to $n(\mathbf{r})$ since the OPE equation 
is not of the Gross-Pitaeviskii type. While the order parameter $\psi ({\mathbf r})$
vanishes at the boundaries $R_{n, \pm}$ between the superfluid and Mott shells, 
and reaches the maximum value $\vert \psi ({\mathbf r}) \vert^2_{max} = (n + 1)/4$,
when $\mu_\mathbf{r} = nU$, the average particle density $n( \mathbf{r} )$ interpolates 
harmonically between Mott shells $n + 1$ and $n$, having the 
average value of $n + 1/2$ when $\mu_\mathbf{r} = nU$.
Furthermore, we show in Fig.~\ref{fig:1-2} that when $t = 1.25 \times 10^{-2} U$, and $\Omega = 6 \times 10^{-6} U$, and $\mu = 2.2U$
the central shell is superfluid, and the order parameter has a local minimum at the origin of the harmonic trap. 
As $\mu$ increases, this minimum is reduced to zero at a critical value, and
the $n = 3$ Mott shell emerges. This property of the emergence 
of a Mott shell from the center of the trap by supressing the superfluid order parameter at that region is a generic feature,
and can be inferred directly from the phase diagram of Fig.~\ref{fig:one} 
via the substitution $\mu \to \mu_{\mathbf{r}}$. 

The local compressibility 
\begin{equation}
\label{eqn:compressibility}
\kappa (\mathbf {r})=\frac{\partial n(\mathbf{r})}
{\partial \mu }=\frac{1}{2 z t (n+1)}
\end{equation}
of the superfluid shells is non-zero, 
in contrast to the incompressible ($\kappa = 0$) $n$ and $n + 1$ Mott shells
for  $r < R_{n, -}$ and $r > R_{n, +}$, respectively. The local compressibility 
indicates the presence of large (small) excitation energies
in the Mott (superfluid) regions, as discussed later. 

We note that only near the edges of the superfluid region
(where $r\approx R_{n, \pm}$ and  $\psi(\mathbf{r})\approx 0$) 
a direct expansion of the OPE (Eq.~\ref{eqn:order-parameter}) 
leads to the effective Gross-Pitaeviskii equation
\begin{equation}
\label{eqn:Gross-Pitaeviskii}
\left( -\frac{1}{2m_{\rm eff}}\nabla^{2}
+ V_{\rm eff} (\mathbf{r}) + 
g_{\rm eff} \left|\psi(\mathbf{r})\right|^{2} \right) \psi(\mathbf{r})= 0.
\end{equation}
Here, $\hbar = 1$,  $m_{\rm eff} = 1/2a^{2} t$ is exactly the boson band mass due to the optical lattice, 
$V_{\rm eff} (\mathbf{r}) =  |n U - \mu_{\mathbf{r}}|/(n + 1) - zt$ is 
the effective potential, and
$g_{\rm eff} = 2zt/(n + 1)$ is the effective interaction.
Notice that $V_{\rm eff} (\mathbf{r}) \le 0$ and vanishes at the boundaries
$R_{n,\pm}$ of the superfluid region since $|n U - \mu_{\mathbf{r}}| =  zt(n+1)$ there. 
Furthermore, $g_{\rm eff} = zt/(n + 1)$ is small in comparison to $U$, indicating
that the superfluid near the edges is weakly interacting, and more so as the Mott index $n$ increases.
When $t \to 0$ $(m_{\rm eff} \to \infty)$ then $\left|\psi(\mathbf{r})\right|^{2} = -V_{\rm eff}/g_{\rm eff}$
leading to the correct limiting behavior of Eq.~\ref{eqn:order-parameter} near $r\approx R_{n, \pm}$.

As discussed above, our nearly degenerate perturbation theory method provides a good description of 
the emergence of superfluid shells. This method is a generalization of the perturbation theory 
developed in~\cite{stoof-2001, fisher-1989} for the uniform case, which was believed not to be extendable to describe
the emergence of superfluid in harmonically confined optical lattices~\cite{barankov-2007}. 
Our method, described above and in~\cite{mitra-2007}, which relies on direct diagonalization of an 
effective two level system, thus provides the connection between the 
mean-field pseudo-spin picture described in~\cite{barankov-2007}, and the perturbative 
approach described in~\cite{stoof-2001}. 

Next, to clarify when the standard theory 
developed in~\cite{stoof-2001, fisher-1989} breaks down, we compare 
the results from our nearly-degenerate perturbation theory with 
the non-degenerate case.

\subsection{Non-degenerate perturbation theory}

When the local energies for Mott phase $n$ with energy 
$E_{0,n} (\mathbf{r}) = U n(n-1)/2 - \mu_{\mathbf{r}} n$,
and Mott phase $n + 1$ with energy
$E_{0,n + 1} (\mathbf{r}) = U(n + 1)n/2 - \mu_{\mathbf{r}} (n + 1)$,
are away for the degeneracy region $\mu_{\mathbf{r}} = nU$, then the
correction to the local energy $E_{0, n}$ is 
\begin{equation}
\label{eqn:energy-second-order}
E_n^{(2)}= \sum_{m \neq n} 
\frac{|\langle n|V|m \rangle|^2} { E_{0,n} - E_{0,m} },
\end{equation}
where $|m \rangle$ denotes the unperturbed wave function of the excited state with eigenvalue $E_{0,m}$.
Here, $V = - c_{\mathbf{r}}\Delta^*(\mathbf{r}) - c_{\mathbf{r}}^{\dagger} \Delta (\mathbf{r})$ 
couples only to states with one more or
one less atom than in the ground-state, and represents  
the perturbation to the Hamiltonian $H_{0,n} (\mathbf{r})$ defined in Eq.~\ref{eqn:interaction-local}.
The, fourth order correction can also be calculated using higher-order perturbation theory~\cite{stoof-2001},
and leads to the Ginzburg-Landau energy
\begin{equation}
\label{eqn:GL-energy-delta}
E_{n} = a_0 + \Lambda (\mathbf{r}) +  a_2 \vert \Delta (\mathbf{r}) \vert^2 + 
a_4 \vert \Delta (\mathbf{r}) \vert^4,
\end{equation}
The coefficients $a_0$, $a_2$ and $a_4$ are all functions of
parameters $n,U,\mu_{\mathbf r},$ and $t$. 
The term $a_0$ corresponds to the unperturbed energy $E_{0,n} (\mathbf{r})$ described 
in Eq.~\ref{eqn:local-energy} and associated with the unperturbed Hamiltonian $H_{0,n}$ defined
in Eq.~\ref{eqn:interaction-local}, while $\Lambda(\mathbf{r})$ is the energy shift shown in 
Eqs.~\ref{eqn:bose-superfluid-local} and~\ref{eqn:lambda}, which contains spatial derivatives
of the order parameter field $\psi (\mathbf{r})$.

The second order coefficient
\begin{equation}
a_2
=\frac{n}{U(n-1)-\mu_{\mathbf r}}+
\frac{n+1}{\mu_{\mathbf r} - U n} + \frac{1}{zt},
\end{equation}
determines the existence of superfluid regions,
when $a_2 < 0$, while the fourth order coefficient
\begin{eqnarray}
\label{eqn:coefficient-fourthorder}
a_4 
&=&\frac{n(n-1)}{\left[ U (n-1)-\mu_{\mathbf r}\right]^2
                 \left[U (2n-3) - 2\mu_{\mathbf r}\right]
                }
                \nonumber \\
&+&   \frac{(n_{\mathbf r}+1)(n+2)}{\left[\mu- U n \right]^2
                \left[2\mu_{\mathbf r}- U(2(n+1))\right]} 
                \nonumber \\
        & - &   \left(\frac{n}{U (n-1)-\mu_{\mathbf r}}+
                \frac{n+1}{\mu_{\mathbf r}- U n} \right) \nonumber \\
        &\times&\left(
                \frac{n}{\left(U (n-1)-\mu_{\mathbf r}\right)^2}+   
                \frac{n+1}{\left(\mu_{\mathbf r}-U n\right)^2}\right) 
\end{eqnarray}
is always positive and is essentially identical to the homogeneous non-degenerate limit~\cite{stoof-2001} except
for the replacement of $\mu \to \mu_{\mathbf{r}}$. However, the inherent inhomogeneity
of the trap potential manifest itself in the local energy for the superfluid regions 
(Eq.~\ref{eqn:GL-energy-delta}) through the spatial dependence of the coefficients
$a_0, a_2$, and $a_n$ and through the spatial derivatives of $\psi (\mathbf{r})$ contained
in $\Lambda (\mathbf{r})$. Thus, we define the local Ginzbug-Landau energy difference
$\Delta E_n = E_n - a_0$, which, in terms of the order parameter $\psi$, becomes
\begin{equation}
\label{eqn:GL-energy-psi}
\Delta E_n  = - ta^2 \psi^* \nabla^2 \psi 
+ a_2 z^2t^2\left|\psi(\mathbf{r})\right|^2
+ a_4 z^4 t^4 \left| \psi\right(\mathbf{r})|^4.
\end{equation}
Here, we retained only terms up fourth order in $\psi$, and up to 
second order in derivatives of $\psi$. 

Minimizing $\Delta E_n$ to zeroth order in the spatial derivatives of $\psi$, 
leads to  
\begin{equation}
\left|\psi({\mathbf r})\right|^2 = - \frac{a_2} {2 a_4 z^2 t^2},
\end{equation}
and setting $\left|\psi({\mathbf r})\right| = 0$ (or $a_2 = 0)$
leads to the local chemical potential
\begin{equation}
\label{eqn:local-chemical-potential}
2\mu_{\mathbf r}^{\pm} = U (2n-1) - zt 
\pm \sqrt{U^2 - 2U(2n+1)zt+ z^2t^2},
\end{equation}
which determines the inner and outer radii
of the Mott shell with filling $n$, where the order parameter $\psi (\mathbf{r}) = 0$ 
vanishes. Solving Eq.~\ref{eqn:local-chemical-potential}
gives the smaller radius of the Mott shell with filling $n$,
\begin{equation}
R_{n, +} = R_{c,n} \sqrt{ 1 + \frac{2 zt (n+1)}{\Omega} \frac{a^2}{R_{c,n}^2} },
\end{equation}
and the larger radius
\begin{equation}
R_{n-1, -} = R_{c,n-1} \sqrt{ 1 - \frac{2 zt n}{\Omega} \frac{a^2} {R_{c,n-1}^2 } },
\end{equation}
calculated to order $t^2/U$. Here, $R_{n, +} > R_{c,n}$, 
and $R_{n-1, -} < R_{c, n-1}$, where $R_{c,n}$ radius of the Mott shell with filling
$n$ when $t = 0$, as defined in Eq.~\ref{eqn:mott-shell-radius}.
Notice that the size of the Mott region $\Delta R_{n, Mott} = R_{n-1,-} - R_{n, +}$ always
decreases with increasing $t$ from its value $R_{c,n-1} - R_{c,n}$ at $t = 0$, 
showing that superfluid regions emerge at the expense of shrinking the 
Mott insulator shells.

The corresponding radii defining the superfluid
region between the Mott shells with filling factors $n$ and $n+1$ are 
precisely $R_{n, +}$ and $R_{n, -}$, 
where $R_{n, \pm}$ are defined in Eq.~\ref{eqn:radii}.
Thus, to order $t^2/U^2$, the thickness of the superfluid shells
$\Delta R_n$ is again given by Eq.~\ref{eqn:width-superfluid},
and we recover the results obtained from our 
degenerate perturbation theory. 

Furthermore, minimization of $\Delta E_n$ (Eq.~\ref{eqn:GL-energy-psi}) 
with respect to $\psi$ reduces to the same Gross-Pitaeviskii equation
described in Eq.~\ref{eqn:Gross-Pitaeviskii} 
near the boundaries $R_{n,+}$ and $R_{n,-}$,
where the order parameter $\psi (\mathbf{r})$ vanishes.
The mapping near the boundaries is $a^2 t \to 1/2 m_{\rm eff}$,
$a_2 z^2 t^2 \to V_{\rm eff} (\mathbf{r})$, and 
$a_4 z^4 t^4 \to g_{\rm eff}$.

\begin{figure}
\scalebox{0.60}{\includegraphics{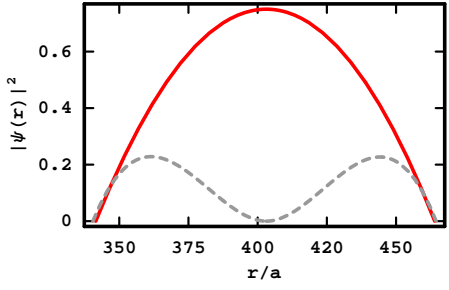} } 
\caption{\label{fig:ope-perturb} (Color online) 
The squared amplitude of the superfluid 
order parameter $\vert \psi (\mathbf{r}) \vert^2$ is shown 
as a solid line (red) for the nearly degenerate case,
and as a dashed line (gray) for the non-degenerate case.
Notice that near the boundaries where $\psi (\mathbf {r}) \approx 0$,
both methods agree and describe the superfluid accurately. 
However, at the center of the superfluid shell, the
non-degenerate method breaks down.
The parameters are same as in Fig.~\ref{fig:1}, and we show the order parameter for the 
inner most superfluid shell, between $n=2$
and $n=3$ Mott shells.
}
\end{figure}

In Fig.~\ref{fig:ope-perturb} we compare the results of the
non-degenerate perturbation theory with that of our nearly degenerate
perturbation theory for small values of $t/U$. The parameters used
are same as in  Fig.~\ref{fig:1} and we show the results only for the innermost
superfluid shell, between $n=2$
and $n=3$ Mott shells. Notice that the non-degenerate perturbation theory is correct
only very close to the boundaries between the superfluid and Mott regions
where $\psi (\mathbf{r})$ is small, but fails to describe the superfluid phase 
at the center of the superfluid shell corresponding to the degeneracy region.

Having, discussed the non-degenerate perturbation approach and
its breakdown, we analyze next the excitation spectrum in the
Mott and superfluids regions.

\section{Excitations in Mott and Superfluid regions} 
\label{sec:excitations}

In this section, we discuss relevant excitations in the Mott and superfluid
regions, and we use the method of functional integrals to obtain quasiparticle
and quasihole excitations in the Mott shells and sound and vortex excitations
in the superfluid shells.

\subsection{Quasiparticle and quasihole excitations \\ in the Mott regions}

The excitation spectrum
in the Mott shells can be obtained using 
the functional integration method~\cite{stoof-2001}
for the action ($\hbar = k_B = 1$, $\beta = 1/T$)
\begin{equation}
S[c^\dagger,c] = 
\int_{0}^{\beta}d\tau\sum_{\mathbf{r}}
\left[ 
c_{\mathbf{r},\tau}^{\dagger} {\partial_\tau} c_{\mathbf{r}, \tau}\nonumber  + H 
\right]
\end{equation}
\begin{figure} [htb]
\scalebox{0.70} {\includegraphics{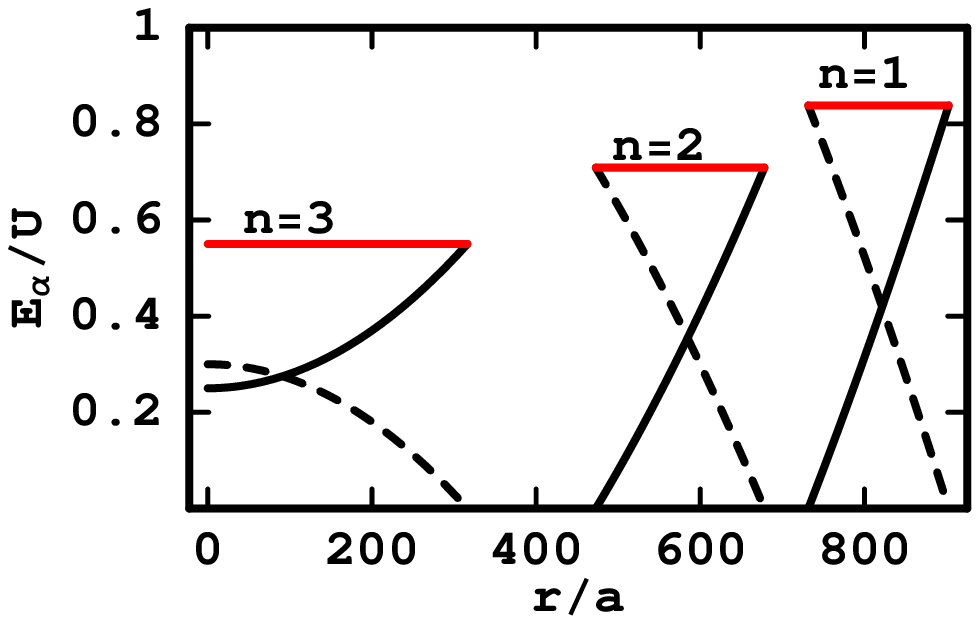}} 
\scalebox{0.70} {\includegraphics{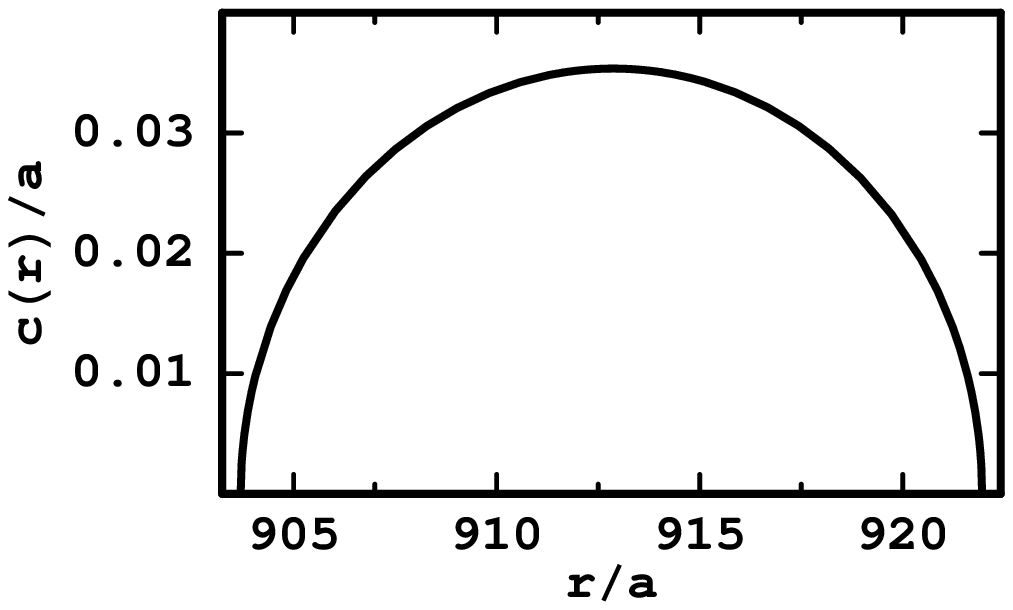}} 
\caption{\label{fig:2} (Color online)
a) Quasiparticle $E_{qp}$ (solid line), quasihole $E_{qh}$ (dashed line),and Mott gap $E_g$ (red) 
energies for $\mathbf{k}= 0$ versus $r/a$.  
b) Sound velocity for the outermost superfluid ring versus $r/a$. 
Same parameters as in Fig.~\ref{fig:1}}.
\end{figure}
leading to the partition function 
$Z = \int {\cal D} c^\dagger {\cal D} c \exp{\left(- S[c^{\dagger},c]\right)}$.  
In each Mott shell we introduce a Hubbard-Stratonovich field
$\Psi$ to take into account fluctuations due to the presence of finite hopping, and integrate out
the bosons $(c^\dagger,c)$ leading to an effective action 
$$
S_{\rm eff}[\Psi^{\dagger},\Psi] = 
\int d\mathbf{r}\sum_{i\omega,\mathbf{kk'}}
\Psi_{i\omega,\mathbf {k}} \Psi_{i\omega,\mathbf {k'}}^{*}  e^{ {i(\mathbf{k}-\mathbf{k'})\cdot\mathbf{r}} }
G_\mathbf{kk'}^{-1} (i\omega, \mathbf r)
$$
to quadratic order in $\Psi^{\dagger}$ and $\Psi$, where
$\mathbf{k}$, $\mathbf{k'}$ are momentum labels, $\omega$ are Matsubara frequencies,
and 
$$
G_\mathbf{kk'}^{-1} (i\omega, \mathbf r) = 
\epsilon_{\mathbf{k'}} 
\left[ 1 + 
\epsilon_\mathbf{k} \left( \frac{n+1}
{i\omega - E_1 (\mathbf{r})}  - \frac{n}{i\omega - E_2 (\mathbf{r})} \right) \right], 
$$ 
with $E_1 (\mathbf{r}) = nU - \mu_{\mathbf{r}}$, $E_2 (\mathbf{r}) = (n-1)U -\mu_{\mathbf{r}}$,
and 
The poles of $G_\mathbf{kk'} (i\omega, \mathbf r)$ are found upon the analytical continuation
$i \omega = \omega + i\delta$, leading to the local excitation energies
$$
\omega_{\pm}=
-\mu_{\mathbf{r}}+\frac{U}{2}(2n-1)-\frac{\epsilon_{\mathbf{k}}}{2}
\pm\frac{1}{2}\sqrt{\epsilon_{\mathbf{k}}^{2}-(4n+2)\epsilon_{\mathbf{k}}U + U^{2}}
$$
where the $+$$(-)$ sign labels quasiparticle (quasihole) excitations.
The energy to add a quasiparticle is $ E_{qp} = \omega_+ $, while 
the energy to add a quasihole is $ E_{qh} = -\omega_- $.

Figure~\ref{fig:2}a shows the quasiparticle and quasihole energies
of the Mott phase as a function of position $\mathbf{r}$. The energy cost 
to create a quasiparticle (quasihole) is minimum (maximum) at the trap
center and increases (decreases) radially, while the Mott-Hubbard gap 
$E_{g} = {\rm min}_{\bf k} ( E_{qp} + E_{qh} )$
is large and independent of $\mathbf{r}$. Thus, 
$ 
E_{g} = {\rm min}_{\bf k} 
\sqrt{ 
\epsilon_{\mathbf{k}}^2 - 4(n + 2) \epsilon_{\mathbf{k}} U + U^2 
}, 
$
which leads to the final result
$ 
E_{g} = 
\sqrt{ (zt)^2 - 4(n + 2) zt U + U^2}. 
$
This expression indicates that it is easier to create a quasiparticle-quasihole excitation
inside higher $n$ Mott shells. The horizontal solid (red) line 
in Fig.~\ref{fig:2}a also shows this tendency. Since this gap is reduced with increasing
$n$ the Mott-insulator gets weaker and thus more susceptible to superfluid fluctuations.
Notice that $E_g$ vanishes when $zt/U = (2n + 1) \pm 2\sqrt{n(n+1)}$, but the physical
solution for the critical value of $zt/U$ corresponds to the largest value, and thus
the plus $(+)$ sign.

The large excitation energies in the Mott regions give away to low energy
excitations in the superfluid regions as discussed next.

\subsection{Excitations in superfluid regions} 

While the creation of quasiparticle-quasihole excitations in the Mott regions
can be energetically costly, the creation of single quasiparticle excitations,
sound waves and vortex excitations in the superfluid shells are more easily 
accessible.

{\it Single quasiparticle excitations:}
The single quasiparticle excitation energy can be
read off from Eq.~\ref{eqn:eigenvalues}, since the first
excited state of $H_{\bf r}^{\rm eff}$ is $E_{+} (\mathbf{r})$,
and the ground state is $E_{-} (\mathbf{r})$.
The local energy difference is $\Delta E = E_{+} (\mathbf{r}) - E_{-} (\mathbf{r})$
is independant of  $\mathbf{r}$. This can be seen from the expression
$\Delta E = 2 \sqrt{ \vert E_d (\mathbf{r}) \vert^2
+ (n+1)\left|\Delta(\mathbf{r})\right|^{2}}$, when the approximate result 
$\Delta(\mathbf{r}) \approx tz \psi (\mathbf{r})$ is used in combination with
Eq.~\ref{eqn:order-parameter-saddle} and with the definition
$E_d(\mathbf{r}) = 
\left[ E_{0, n+1}(\mathbf{r})-  E_{0, n}(\mathbf{r})\right]/2 = ( nU - \mu_{\mathbf{r}} )/2$.
The final answer is $\Delta E \approx (n+1) t z$,
which indicates that energy cost of adding a quasiparticle in the superfluid
shell is small in comparison to the cost of adding a quasiparticle in the Mott
region. 

It is also interesting to analyze the eigenvectors of the local Hamiltonian
defined in Eq.~\ref{eqn:eigenvalues}. The eigenvector corresponding to $E_+$ is
\begin{equation}
\label{eqn:eigenvectors1}
\vert E_{+} \rangle = 
\left( 
\begin{array}{c}
\frac{-E_d+\Delta E/2}{\sqrt{n+1}\Delta^*(r)} \\
1
\end{array}
\right) 
\end{equation}
and reduces to the vector $(0,1)^{T}$ corresponding to the energy $E_n$ of the Mott phase with 
filling $n$, when $\Delta\to 0$. The eigenvector corresponding to $E_-$ is
\begin{equation}
\label{eqn:eigenvectors2}
\vert E_{-} \rangle = 
\left( 
\begin{array}{c}
1 \\
\frac{\sqrt{n+1}\Delta^*(r)}{-E_d-\Delta E/2}
\end{array}
\right) 
\end{equation}
and reduces to the vector $(1,0)^T$  corresponding to the energy $E_{n+1}$ of the Mott phase with filling $n+1$, 
when $\Delta\to 0$.

However, the most interesting excitations in the superfluid regions are 
collective in nature. We will discuss next the collective sound excitations and later the 
appearance of vortices.

{\it Sound velocity:} The excitation spectrum of collective modes in the superfluid region can
also be calculated using the functional integral method.
First we introduce the Hubbard-Stratonovich field $\psi$ which now corresponds to the
order parameter in the superfluid region. Second we use an amplitude-phase representation
$\psi (\mathbf{r}, \tau) =  |\psi (\mathbf{r}, \tau)| \exp{[i\varphi(\mathbf{r}, \tau)]}$
and apply the nearly degenerate perturbation theory described earlier to integrate out
the boson fields $c^\dagger$ and $c$. Thus, we obtain 
the phase-only effective action 
\begin{equation}
\label{eqn:action-superfluid}
S_{\rm eff} = \frac{1}{2 L^d} \int d\mathbf{r}d\tau 
\left[ 
\kappa (\partial_{\tau} \varphi)^2 + \rho_{ij} \partial_i \varphi \partial_j \varphi
\right]
\end{equation}
to quadratic order in the phase variable for the superfluid region between the $n$ and $n+1$ Mott shells. 
Here, we assumed that $|\psi (\mathbf{r}, \tau)|$ is $\tau$-independent
at the saddle point. The coefficient $\kappa$ is the compressibility of the superfluid 
described in Eq.~\ref{eqn:compressibility}, and the local superfluid density tensor
\begin{equation}
\label{eqn:superfluid-density}
\rho_{ij} =
\frac{(n + 1) t a^{2}}{2} \frac {F (\vert \psi \vert, n, t)} {G  (\vert \psi \vert, n, t) }
-2 t a^{2}\left|\psi\right|^{2}\delta_{ij}, 
\end{equation}
where the numerator of the first term is 
$$
F (\vert \psi \vert, n, t) = 
t
\left( 
4z|\psi|^{2}\delta_{ij}
+ 4|\psi|\nabla^{2} \vert \psi \vert \delta_{ij} - 2 \partial_{i}|\psi| \partial_{j}|\psi|
\right)
$$
and the denominator of the first term is
$$
G (\vert \psi \vert, n, t) = 
\sqrt{ 
( nU - \mu_{\mathbf{r}} )^2 / 4 
+ (n+1)t^2 \Gamma (\vert \psi \vert)
},
$$
with the function
$$
\Gamma (\vert \psi \vert) = 
z^2 \vert \psi \vert^2 + 2 \vert \psi \vert \nabla \vert \psi \vert + ( \nabla^2 \vert \psi \vert)^2. 
$$
The complex structure of the superfluid density tensor is a direct consequence of the non-Gross-Pitaeviskii
nature of the order parameter equation (Eq.~\ref{eqn:order-parameter}).

Insight can be gained into the structure of the local superfluid density tensor by neglecting the gradient
terms involving $\vert \psi \vert$, which in combination with Eq.~\ref{eqn:order-parameter-saddle}
produces a local superfluid density tensor
\begin{equation}
\rho_s({\bf r})=\rho_{ii}=2 t a^2 \vert \psi ({\bf r}) \vert^2,
\end{equation}
which vanishes at the Mott boundaries $R_{n,\pm}$. 
This local superfluid density tensor has been described previously in Refs.~\cite{mitra-2007} and~\cite{barankov-2007},
however the more general expression shown in Eq.~\ref{eqn:superfluid-density} goes beyond the
mean field approximation presented in the pseudo-spin description~\cite{barankov-2007}.
In the present approximation, the resulting wave equation has the form
\begin{equation}
\label{eqn:wave-equation}
\partial_\tau^2 \varphi - \partial_i \left[ \frac{\rho_s (\mathbf{r})} {\kappa} \partial_i \varphi \right] = 0, 
\end{equation}
leading to a local sound velocity  
$c ({\mathbf r}) = \sqrt{\rho_s (\mathbf{r})/ \kappa}$, 
which in terms of the order parameter reads
$c ({\mathbf r}) = 2 \sqrt{(n+1)z} t a |\psi (\mathbf{r})|$. The local speed of sound has it
maximal value at $|\psi (\mathbf{r})|_{max} = (\sqrt{n+1})/2$, and the superfluid region behaves
as a medium of continuous index of refraction 
\begin{equation}
\chi  ({\mathbf r}) = \frac{c_{max}}{c ({\mathbf r})} 
= \frac{\sqrt{n+1}}{2 |\psi (\mathbf{r})|}.
\end{equation}
Notice that $\chi  ({\mathbf r}) \to \infty$ at 
the Mott boundaries where $|\psi (\mathbf{r})| = 0$, indicating that the sound waves of the
superfluid do not propagate into the Mott regions. A plot of the local sound velocity is shown 
in Fig.~\ref{fig:2}b for the superfluid region between the $n = 1$ and $n =0$ Mott shells.
From the phase-only effective action for the superfluid region, we can also investigate
the vortex excitations, which are discussed next.

{\it Vortices and antivortices:} Next, we explore vortex solutions 
in two cases where spontaneous vortex-antivortex pairs can appear as indicators of the 
Berezinski-Kosterlitz-Thouless (BKT) transition~\cite{berezinski, kosterlitz-thouless}.
Case I corresponds to a 3D system, where the superfluid regions are very thin 
$\Delta R_n \ll R_{c,n}$, leading to essentially a two-dimensional superfluid in curved space. 
Case II corresponds to a 2D system, where the superfluid regions are 
thick rings $\Delta R_n \sim R_{c,n}$, leading to essentially a two-dimensional superfluid subject
to boundary conditions. 

In a flat space two-dimensional system stationary vortex solutions 
must satisfy $\oint\nabla\varphi\cdot d\mathbf{l}=2\pi m$,
where $m = \pm1,\pm2,...$ is the vorticity (topological charge) and
$\nabla\varphi$ is the superfluid velocity. The standard vortex solution
in cylindrical coordinates is $\nabla{\varphi} = m \hat{\theta}/r$,
and the corresponding free energy per unit volume is
\begin{equation}
{\cal F}= \frac{1}{2 L^d}\int d\mathbf{r}\rho_s(\mathbf{r})(\nabla\varphi)^{2}.
\end{equation}
This situation is analogous to a two-dimensional linear
dielectric material where the displacement field
is 
\begin{equation}
\mathbf{D}=\nabla\varphi\times\hat{z}=\epsilon(\mathbf{r})\mathbf{E},
\end{equation}
with dielectric function $\epsilon(\mathbf{r})=1/\left[ 2\pi\rho(\mathbf{r})\right]$.
Notice that the dieletric function diverges at the superfluid boundaries,
since $\rho_s (\mathbf{r}) \to 0$ in those regions.
In this language ${\cal F}$ is identical to the electrostatic
energy per unit volume 
\begin{equation}
U_{\textrm{el}} = \frac{1}{2 L^d}\int d\mathbf{rD}\cdot\mathbf{E}.
\end{equation}
In general the solutions for several vortices (antivortices) can be obtained from 
\begin{equation}
\nabla \wedge {\nabla \varphi} = 2\pi {\hat \mathbf{z} }\sum_i m_i \delta ({\mathbf r} - \mathbf{r}_i),
\end{equation}
where ${\bf r}_i$ is the location of the vortex (or antivortex) of vorticity $m_{i}$.

\begin{figure} [htb]
\centerline{ \scalebox{0.45} {\includegraphics{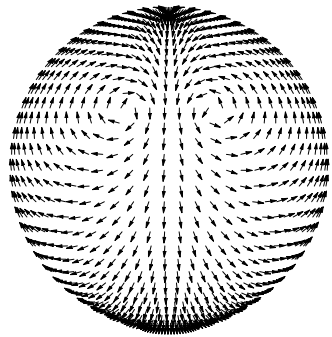}} } 
\caption{\label{fig:3}
Three-dimensional view of a vortex-antivortex pair in 2D superfluid shell separating two 3D Mott regions.
}
\end{figure}

In case I the superfluid state appears below 
$T_{BKT} \approx \pi {\widetilde \rho}_s ( \mathbf{r} = \mathbf{R}_{c,n} )/2$,
where ${\widetilde \rho}_s = \rho_s/a^2$ has dimensions of energy. 
In this limit ${\widetilde \rho}_s ( \mathbf{r} = \mathbf{R}_{c,n} ) = (n+1)t/2$ and
the critical temperature of the superfluid shell between the 
$n$ and $n+1$ Mott regions depends on the index $n$. Notice, however that such estimate
is reasonable only when the radius of the shells are sufficiently large. The
solution for a vortex-antivortex (VA) pair in curved two-dimensional space is  
$$
\label{eqn:vortex-antivortex}
\varphi(r = R_{c,n},\theta,\phi)=
\arctan\left(\frac {4 b R_{c,n}\tan\left(\frac{\pi-\theta}{2}\right)\sin(\phi)}
{b^2 - 4 R_{c,n}^2 \tan^2 \left(\frac{\pi-\theta}{2}\right)}\right)
$$
where $R_{c,n} = a \sqrt{ 2 (\mu -nU)/\Omega}$ is the radius of the superfluid shell,
and $b$ is the VA size.
A three-dimensional view of the velocity field $\nabla \varphi (\theta,\phi)$ is shown
in Fig.~\ref{fig:3}. When the superfluid shell has a small
thickness $\Delta R_n$ then $T_{BKT} \approx  \pi t (n+1) \Delta R_n/6 a $, 
while the vortex-antivortex pair has an approximate 
solution of the same form as above which interpolates between $\varphi( r = R_{n,-}, \theta,\phi)$ and 
$\varphi( r = R_{n,+}, \theta,\phi)$. 

In case II the superfluid regions are rings bounded by $R_{n,-}$ and $R_{n,+}$, and one can 
use the Coulomb gas analogy described above, conformal mapping techniques and proper boundary
conditions to obtain vortex-antivortex solutions.
The creation of vortex-antivortex 
pairs are energetically quite costly when $\Delta R_n \ll R_{c,n}$, due to strong
confinement effects of the boundaries,
thus we do not expect a BKT-type superfluid transition to occur until 
$\Delta R_n$ is substantially large ($\sim R_{c,n}$). 
Only in this limit, we expect a BKT transition with
$T_{BKT} \approx {\pi \langle {\widetilde \rho}_s \rangle/2}$, 
where $\langle {\widetilde \rho}_s \rangle/2$ is the surface area average of 
${\widetilde \rho}_s ({\mathbf{r}})$.

Having analyized the excitation spectra in the insulator and superfluid regions,
we discuss next some possible experimental detection schemes for the superfluid shells.

\section{Detection of Superfluid Shells}
\label{sec:experiments}

Eventhough the use of tomographic microwave techniques has allowed the detection 
of several Mott regions~\cite{MIT-2006, Mainz-2006}, it has been quite challenging
to detect superfluid shells. In this section, we propose a possible experiment 
for the detection of superfluid shells through the use of Gauss-Laguerre and Gaussian beams 
followed by Bragg spectroscopy. The idea is that Gauss-Laguerre and Gaussian beams can tranfer
angular momentum to the atoms in superfluid phase without transfering linear momentum, and that Bragg spectroscopy can detect
their existence since the technique is only sensitive to the velocity of
the atoms in the superfluid phase. This is because the Mott insulator regions do not absorb the angular
momentum because of the presence of a large gap in the excitation spectrum.
Next, when the frequencies of the beams used in Bragg spectroscopy are correctly chosen some of the 
rotating atoms in the superfluid regions acquire extra momentum and are kicked out of their shells and can be imaged.

The Gauss-Laguerre technique has been sucessfully used to rotate 
superfluid sodium atoms $^{23}$Na in a {\it mexican hat} potential~\cite{kris-2006, kris-2007}.
The technique used was a two step process, where initially a 
Gauss-Laguerre beam transfered both linear 
and angular momentum to atoms, and a second
Gaussian beam canceled the linear momentum transfer leaving each trapped 
atom with exactly one unit of angular momentum. In this manner, the atoms participating 
in superfluidity rotate without dissipation, and when the trap is released, the superfluid region
expands, but does not fill the center of the cloud, thus maintaining a toroidal structure 
throughout the time of flight. Furthermore, Bragg spectroscopy was used to detect the sense
of rotation of vortices within a vortex lattice of sodium BEC~\cite{muniz-2006} and
to observe the persistent flow of Bose-condensed Sodium atoms in toroidal traps~\cite{kris-private-2007}.

The Gauss-Laguerre technique and Bragg spectroscopy may also be used to detect 
superfluid shells of bosons in harmonically confined optical lattices. However, the situation 
here is slightly different because of the existence of Mott shells and multiple superfluid regions. To illustrate this we discuss 
the simpler case of a nearly two-dimensional configuration, where the harmonic trap is 
very tight along the z-direction and loose along the x- and y- directions and only two
superfluid shells are present. Upon application of the Gauss-Laguerre technique along the 
z- direction, angular momentum transfer occurs essentially to the atoms in the 
superfluid phase, imposing a rotating superfluid current with a well defined superfluid
velocity profile, while the Mott regions remain unchanged due to their large gap in the excitation 
spectrum. The angular momentum transfer is chosen to occur along the z direction
generating the superfluid currents in the x-y plane, and the amount of angular momentum
transfer is assumed to be $\hbar$ for each atom that absorbed a Gauss-Laguerre photon with $l=1$. 
Next we use Bragg spectroscopy with counter-propagating beams along the x-direction to detect the sense of rotation
and determine the regions of superfluid rings with well defined velocities. If there 
is sufficient optical resolution and signal-to-noise ratio, then the experiment may be performed
{\it in situ}, otherwise the Bragg spectroscopic measurements can be performed in time-of-flight.
\begin{figure} [htb]
\centerline{ \scalebox{0.65} {\includegraphics{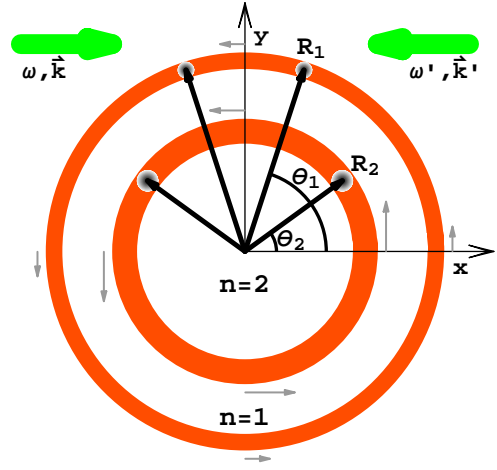}} } 
\caption{\label{fig:brag}
(Color online) Schematic plot for the detection of superfluid shells using Bragg spectroscopy.
The angles $\theta_1$ and $\theta_2$ indicate the locations
of strongest momentum transfer from the Bragg beams (large green arrows) to the rotating superfluid 
shells of radii $R_1$ and $R_2$. The gray arrows indicate the sense of rotation of the superfluid shells.
}
\end{figure}

A schematic plot of the detection of superfluid shells using Bragg spectroscopy can be found 
in Fig.~\ref{fig:brag}, which shows two superfluid shells rotating counter-clockwise and 
two counter propagating beams applied along the x direction. As can be seen in Fig.~\ref{fig:brag} the right(left)-going 
beam has frequency $\omega$($\omega'$) and linear momentum ${\bf k}$(${\bf k'}$). The bose atoms undergo a transition from the
internal state with energy $\epsilon_i$ to the internal state $\epsilon_f$. In the following analysis, 
we retain $\hbar$, instead of setting $\hbar = 1$. Applying momentum conservation, 
we can easily obtain that the final linear momentum of the atoms in the superfluid region 
is ${\bf p}_f={\bf p}_i+\hbar(k+k')\hat{\bf x}$, in terms of the initial 
linear momentum of the photons $\hbar k\hat{\bf x}$ and -$\hbar k'\hat{\bf x}$. Thus, the Bragg
beams transfer a net linear momentum $\hbar(k+k'){\bf x}$ to the
atoms which satisfies the energy conservation condition
\begin{equation}
\label{eq:bragg}
\hbar(\omega-\omega') = \epsilon_f - \epsilon_i - v_x\hbar(k+k') + \frac{\hbar^2 (k+k')^2}{2m}
\end{equation}

In the equation above $v_x$ is the component of the velocity ${\bf v}_i = {\bf p}_i/m$ along the x direction. 
Notice that $v_x$ can also be written as $v_x = v_i \sin\theta$ where $\theta$ is 
the angle between the Bragg beams and the velocity ${\bf v}_i$ of the atoms in the superfluid shell 
and $v_i=p_i/m$ is the magnitude of ${\bf v}_i$. 
For an atom carrying one unit of  angular momentum, the superfluid velocity is ${\bf v}_i = \hbar\hat{\theta}/mr$. Therefore,
within a superfluid shell at position $r=R$ atoms get a linear momentum kick of $\hbar(k+k') \hat{\bf x}$ when the velocity  
$v_x=\hbar\sin\theta/mR$ satisfies the condition given in Eq.~\ref{eq:bragg}. 
This leads to two Bragg angles $\theta=-\sin^{-1}(mRv_x/\hbar)$, and $\pi-\theta$ for each 
superfluid shell. As can be seen in Fig.~\ref{fig:brag} the Bragg angles are $\theta_1$ and $\pi-\theta_1$ 
for the outer superfluid shell
labelled by  $R_1$, and are $\theta_2$ and $\pi-\theta_2$ for the inner superfluid shell
labelled by $R_2$. The regions with same velocity $v_x$ are identified by the condition 
$\sin\theta_1/R_1=\sin\theta_2/R_2=mv_x/\hbar$.
Once these atoms are kicked out of their respective superfluid shells, they form two small cloud, which can be 
detected by direct imaging. As mentioned before, 
if there  is sufficient optical resolution and signal-to-noise ratio, then the experiment may be performed
{\it in situ}, otherwise the Bragg spectroscopic measurements can be performed in time-of-flight. In fact
these clouds also carry the information of the sense of rotation of the superfluid shells and their velocity profile, 
and one can in principal 
extract that information from the images. 
    
Having discussed our proposal for the experimental detection of superfluid
shells in harmonically confined optical lattices, we present next our conclusions.

\section{Conclusions} 
\label{sec:conclusions}

We studied 2D and 3D optical lattices of atomic or molecular bosons in harmonically confining
potentials, and showed that between the Mott regions of filling fraction $n$ and $n+1$,
superfluid shells emerge as a result of fluctuations due to finite hopping.
We found that the presence of finite hopping breaks 
the local energy degeneracy of neighboring Mott-shells, determines
the size of the superfluid regions as shown in Fig~\ref{fig:1}, 
and is responsible for low energy (sound) and vortex excitations. 
In addition, we obtained an order parameter equation which is not
in general of the Gross-Pitaeviskii type, except near the boundaries
separating the superfluid from the insulating regions.
Furthermore, we obtained bound vortex-antivortex solutions (as shown in Fig~\ref{fig:3}
below the Berezinski-Kosterlitz-Thouless (BKT) transition
when superfluid regions are thin (nearly 2D) spherical (or ellipsoidal) shells.
Finally, we discussed that the emergence of these superfluid regions should be detectable 
using a combination of Gauss-Laguerre and Bragg spectroscopy techniques.

We thank NSF (DMR 0304380, and PHY 0426696) for financial support. We thank F. W. Strauch, 
E. Tiesinga, C. W. Clark, I. Spielman, W. D Phillips, and K. Helmerson for discussions, and 
we thank R. A. Barankov, C. Lannert, and S. Vishveshwara for drawing our attention 
to their related work~\cite{barankov-2007}.

\end{document}